\documentclass[12cm]{article}
\usepackage{amsmath}
\usepackage{graphicx}
\usepackage{caption}
\usepackage{subcaption}
\usepackage{euscript}
\usepackage{amssymb}
\newcommand{\nn}{\nonumber}
\usepackage{authblk}
\usepackage{appendix}
\DeclareCaptionSubType{figure}
\title{Characterization and reduction of variability in selection based on effect-size using association measures in cohort study of heterogeneous diseases.}
\author[1]{Venkateshan Kannan}  
\author[2]{Kristina Alexandersson}
\author[1]{Jesper Tegner}
\affil[1]{Computational Medicine Unit, Department of Medicine, Solna, Karolinska Institutet, SE-17176, Stockholm, Sweden}
\affil[2]{Division of Insurance Medicine, Department of Clinical Neuroscience, Karolinska Institutet, SE-17177 Stockholm, Sweden}
\date{}
\begin{document}
\maketitle

\begin{abstract}
Cohort studies employ pairwise measures of association to quantify dependencies among conditions and exposures. To reliably use these measures to draw conclusions about the underlying association strengths requires that the measures be robust and unbiased. These considerations assume greater significance when applied to disease networks, where associations among heterogeneous pairs of diseases are ranked. Using disease diagnoses data from a large cohort of 5.5 million individuals, we develop a comprehensive methodology to characterize the bias of standard association measures like relative risk and $\phi$ correlation.  To overcome these biases, we devise a novel measure based on a stochastic model for disease development. The new measure is demonstrated to have the least overall bias and hence would be most suitable for application to heterogeneous disease cohorts.
\end{abstract}

\maketitle

\section{Introduction}
\label{s:intro}

Population cohort is used in epidemiological and medical research to infer important relationships between diseases and related factors.  These relationships are deduced based on statistical measures \cite{Breslow1980,Greenland1988}  employed to quantify the nature and degree of association. \medskip 

The validity of inferences made using such measures therefore depends on the consistency and robustness of the measure. In qualitative terms, a pairwise association measure should represent the `true interaction' between a pair of diseases and this should be correctly identified for a wide range of disease pairs. We say that a measure has a bias when there is a systematic dependence of its evaluation on some characteristic of the disease pair, such as the prevalence rate of eiher disease. These issues acquire an even greater significance when considering comparison and rankings of disease pais of a heterogeneous dataset as in the case of disease networks \cite{Hidalgo2009,Chmiel2014,Kannan2016}.  \medskip

The straightforward approach to investigate this potential problem would be to simulate several pairs of diseases with a similar `interaction' but differing with respect to the other quantities (those independent of joint co-occurrence) and determining if there is a systematic dependence of the measure on these quantities. The difficulty with this approach is that there is no objective definition of `interaction' of a pair of diseases; indeed, the measure is trying to precisely capture this, leading to circularity in the problem statement. And yet, this phenomena does not arise from an artifice in semantics; its origins lie in our lack of understanding of how a pair of related diseases interact. \medskip


In cross-sectional studies, where statistical associations are only cautiously interpreted as potential causal relations \cite{Grimes2002}, this lack of understanding is accepted. Nonetheless, what is less appreciated is how similar values of association obtained by applying a single measure may reflect different extents of true association, even when other sources of bias such as selection type or confounder effect is absent. \medskip  



Although there have been studies describing the differences in measure properties and suggested procedures for domain-specific selection \cite{Tan2002,Mosteller1968}, to the best of our knowledge, there has been no systematic approach to examine the bias in association measures in the context of diseases. The basis of such an approach would be a large cohort data where measure bias would be empirically determined. Importantly, the difficulty of accurate simulation of disease associations discussed earlier is directly overcome with a dataset consisting of prevalence and co-occurrence of a heterogeneous set of diseases. This enables us to define bias by systematic over or under estimation of associations depending on underlying putative bias variables. This also provides an unambiguous method to compare the bias of different measures. \medskip

In this work, we use recorded diagnoses for a cohort $\mathfrak{R}$ of all 5.5 million people living in Sweden  followed prospectively for 13 years. We develop a systematic methodology to characterize the biases of standard measures of associations between disease pairs in this cohort.  With the aim of understanding the impact of bias on effect-size calculations, we quantify the bias using three distinct indices. Our analysis conclusively demonstrates the prevalence and nature of measure bias in cross-sectional studies and provides insights into its origins. \medskip

To address these biases, we derive, from first principles, a novel measure family using  stochastic model of disease development with differential rate of diagnosis.  The new measure shows significantly better performance uniformly across a wide range of parameter considerations and bias variables. Overall we find the least bias with the new measure among all those that were considered. We repeat the same analysis with the U.S. Medicare data of 32 million individuals and once again, find that our new measure outperforms all others.  We thus provide theoretical justification for its proposal using the stochastic model and empirical demonstration of its characteristics using two large datasets.\medskip
\section{Measure Bias}
\subsection{Association measures} 
\label{s:SMA} 
We introduce the framework for characterizing bias by considering two commonly used  measures for pairwise associations, relative risk (RR) \cite{Pearce2012, Wolf1991} and $\phi-$correlation \cite{Kramer2012, Wang2012}. RR and the family of measures similar to it such as odds ratio \cite{Zhang1998, Szumilas2010}, hazard ratio, and Yule's Q \cite{YULE1903} are based on relative probabilities of occurrence of diseases in different conditions. To consider these measures in more general terms, we define the $2 \times 2$ contingency table (Table \ref{Cont}) for a disease pair $A$ and $B$, where the top-left entry $p$ represents the number of individuals having both diseases, $q$, the number having  $B$ but not $A$ and similarly for the entries on the second row.  $N= p +q +r +s$ is the total number of individuals. \medskip
  
The relative risk for disease A, in the presence of $B$ is the ratio of prevalence of $A$ in the subset diagnosed with $B$ to the prevalence in the subset not diagnosed with $B$.  :
\begin{equation}
RR_{A|B} =  \frac{p/(p+q)}{r/(r+s)} 
\end{equation}     
and likewise, swapping $A$ and $B$
 $$RR_{B|A} =  \frac{p/(p+r)}{q/(q+s)} $$
 
It should be noted that the original definition of relative risk considers disease incidence among sets exposed and not exposed to a given condition. This explains the asymmetry in the above definition, where presence of a disease is considered as an exposure condition. However, we cannot apply the original interpretation directly to the cohort when the order of occurrence of the disease pair in an individual is unknown. Despite that, we can make simplifying assumptions for large cohort, and this is precisely the definition used in previous studies involving disease networks \cite{Hidalgo2009,Wang2012}.  As the individual disease prevalences in our cohort is small, we can assume that $$p/s,q/s,r/s <<1 .$$ Further if we assume that $$p/r,p/q <<1,$$ which effectively implies that prevalence of one disease within the subset of patients having another disease to also be very small, it is easy to show that the reduced expressions for both are identical. 
\begin{align}
 RR_{ A|B}    & \sim \frac{p N}{n_A n_B}
\label{RR-Symm}          
\end{align}
We find that the expression in the last line above is explicitly symmetric in the two diseases. \medskip

A related measure, odds ratio \cite{Cummings2009} for the same contingency table is given by:
\begin{equation}
 OR = \frac{p/q}{r/s} = \frac{ps}{qr}
\label{OR-Eqn}
\end{equation}

Unlike RR, odds ratio is explicitly symmetric in the two diseases, and further it is easy to show that, in the limit that we are interested in, where prevalence rates are assumed to be small, they converge to the same value.

$$\frac{ps}{qr} = \frac{pN}{n_A n_B} \frac{1}{1+ (p+q+r)/s} (1+p/r)(1+p/q) \sim \frac{pN}{n_A n_B} $$

\vspace{5mm}

{\bf $\phi$ correlation:} \\

Another common measure used for contingency table is the $\phi$ correlation and measures that reduce to a similar form include Cohen's $\kappa$ \cite{Cohen1960} and Kendall's $\tau_b$ \cite{KENDALL1938}.  \medskip

The $\phi$ correlation is obtained by taking the standard correlation between the binary vectors corresponding to the two diseases. For a given disease $A$, the corresponding vector is of length equal to the number of patients and each entry is 1(0) depending on the disease being present (absent) in that individual. 

\begin{equation}
\phi_{A,B} = \frac{\frac{p}{N}- \frac{n_An_B}{N^2} }{ \sqrt{\frac{n_A}{N} \frac{n_B}{N} \left( 1- \frac{n_A}{N}\right) \left( 1- \frac{n_B}{N}\right)}}
\label{Phi-Org}
\end{equation} 
Equivalently, it can be defined as $\sqrt{\chi^2/N}$ where $\chi^2$ is chi-squared statistic calculated for the contingency table. 

%

\subsection{Formulating measure bias}
Let $\mathcal{D}$ be the dataset from a cohort of $N$ individuals and $K$ potential disease diagnoses.  For each individual $p$, we represent their diagnoses using a vector of binary indicator variables $(A^{p}_1,A^{p}_2,\cdots A^{p}_K)$ where $A^{p}_i$ is 1(0) depending on the presence (absence) of disease $i$.  For any arbitrary pair of distinct diseases $(i,j)$, let $n_{uv} = \displaystyle \sum_p \mathbb{1} [ A^{p}_i = u] \mathbb{1} [A^{p}_j = v]$  ($u,v \in \{0,1\}$) where $\mathbb{1}[.]$ is 1(0) if the condition expressed in $[.]$ evaluates to true(false). Define the marginal $n_{u+} = n_{u0} + n_{u1}$ for $u = 0,1$ and likewise for $n_{+v}$. By definition, $\displaystyle \sum_{u= 0,1 ;v = 0,1} n_{uv} = N$. For a given measure $\mu$ let $P(\mu)$ be its distribution across all pairs of diseases in the cohort. We say that the measure has a bias with respect to variable $W$ ($W$ is some characteristic of the disease pair prevalence) if there is a systematic dependence of the measure conditioned on that variable, i.e.,   $P(\mu| W )$ is some non-trivial function of $W$. \medskip

It is important to note that this definition of bias is empirical. The existence or non-existence of bias in a given measure is dependent on the nature of associations found in the dataset. A measure may show considerable bias when applied to a disease cohort but may be well-suited (i.e., show no bias) when measuring association between developmental indices of nations. This is qualitatively different from identifying general properties that a measure is required to satisfy \cite{Olivier2013,Ferguson2009} although it is generally accepted that identifying the property that is most relevant depends on the specific context \cite{Tan2002}. \medskip


Define $\mathcal{F}$ to be the set of distinct diseases (classified by ICD 10 three-character precision) in our cohort $\mathfrak{R}$ with $K = |\mathcal{F}| > 1400$ . We investigate measure bias by considering variables $W$ that are likely to affect the measure characteristics. We first examine potential bias with respect to the expected co-occurrence under the assumption of independence of the diseases, i.e., if  $n_A, n_B$ represent the prevalence of a given disease pair $(A,B)$ then $ W= n_{AB}^{(0)} = N\frac{n_A}{N}\frac{n_B}{N}  = \frac{n_A n_B}{N}$. Observe that there is no {\it a priori} reason to expect the distribution of the measure to depend on it.  \medskip

\begin{figure}[!p]
\hspace{-1cm}
\includegraphics[scale=0.28]{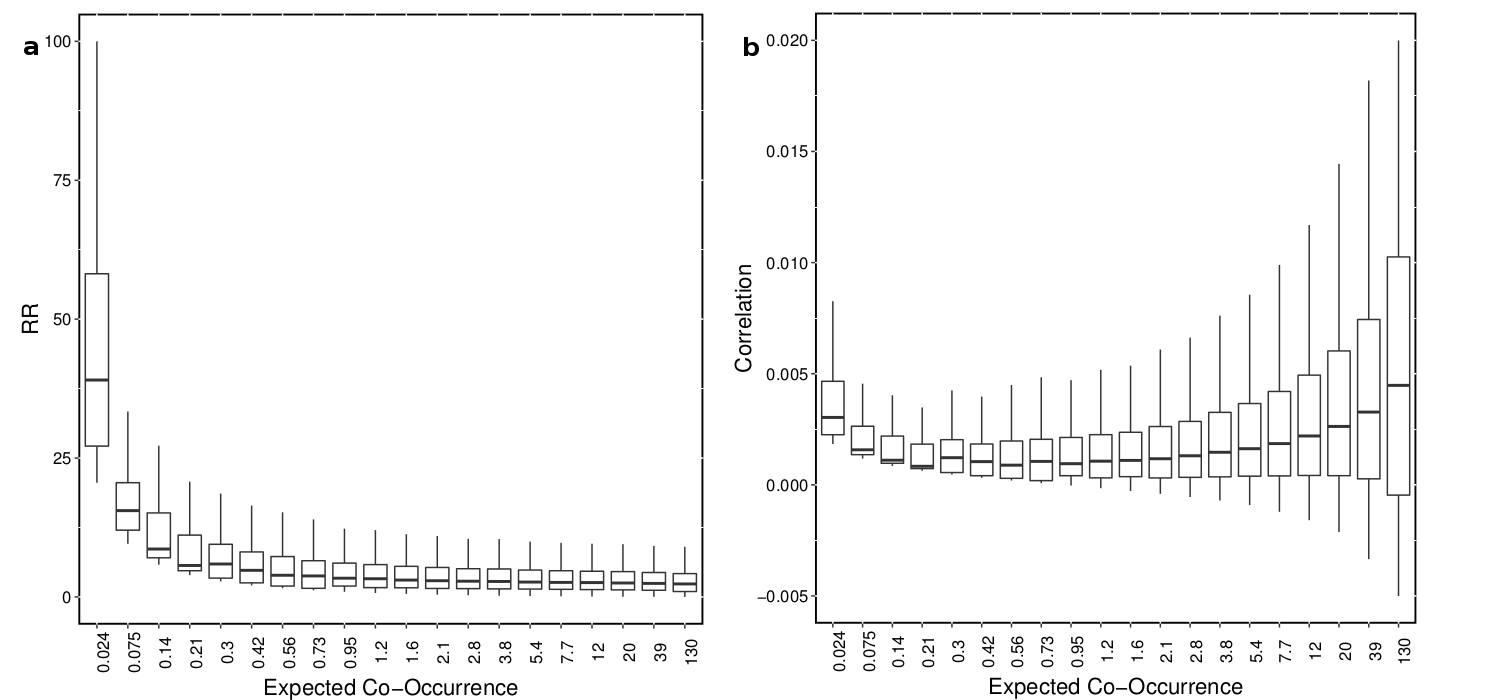}
\caption{(a)Variation of RR as function of the expected co-occurrence shown as a box plot for all the pairs that fill within the particular window of expected co-occurrence. (b) The same analysis repeated for $\phi$ correlation.}
\label{RRCorr-bias}
\end{figure}

In line with our formulation, we partition the set of all disease pairs $\mathcal{D}$ into $M$ (20 in this case, but it could be any reasonable number so that one can get reliable statistics within each interval and at the same time, be able to discern the overall trend across intervals) mutually exclusive subsets $\mathcal{D}_j, j=1,2,\cdots M$. 
 $$ \mathcal{D} = \{  \{ A,B \} | A,B \in \mathcal{F} \}  = \displaystyle \cup_{j=1}^M \mathcal{D}_j $$ 
 where 
$\mathcal{D}_j = \{ \{ A, B  \} |  v_{j-1} < n_{AB}^{(0)} < v_j  \} $ The intervals $v_j$ are determined by the requirement that each of these partitions contain the same number of pairs $| \mathcal{D}_0| =\cdots |\mathcal{D}_j|= \cdots  |\mathcal{D}_M|$. Next we apply the measures within each partition to identify systematic dependence, if any. We begin with RR (Eq. \ref{RR-Symm})  and Fig. \ref{RRCorr-bias}a shows the box plot of the distribution of the measure values within each partition. We find unambiguous systematic bias where low expected co-occurrence leads to higher values of RR. Equally significant is the variation of size of the box representing the boundaries of the 25th and 75th percentile (interquartile range) for the collection of RR values in each partition. This increase is even more pronounced than that of the median, as the expected co-occurrence decreases. \medskip

We can explain the large spread by noting that, for lower expected co-occurrences, small fluctuations in co-occurrence numbers leads to wide variations in RR. Under the assumption of the independence of a disease pair, we can show that the distribution of co-occurrences (for a given disease pair) is Poisson with mean being the expected co-occurrence (see Appendix B). As the variance of a Poisson distribution is equal to the mean, the coefficient of variation is given by $\sqrt{n_{AB}^{(0)}}/n_{AB}^{(0)} =\frac{1}{\sqrt{n_{AB}^{(0)}}}$, representing an inverse relation with expected co-occurrence . \medskip

The same analysis is repeated for the $\phi$ correlation and Fig. \ref{RRCorr-bias}b shows the box plot for this case. A systematic bias is once again immediate from inspecting the figure, except that the bias points in the opposite direction: $\phi$ correlation tends to inflate the associations for higher expected co-occurrences. \medskip
 
We thus find that both RR and $\phi$ correlation have significant bias with respect to the expected co-occurrence, except that the bias works in opposite directions. Where RR tends to assign higher associations to disease pairs with lower expected co-occurrence, the exact opposite it true for the $\phi$ correlation. Conversely, higher expected co-occurrences lead to lower RR and higher $\phi$ correlation. Note that lower (higher) expected co-occurrences arise when one or both diseases have low (high) prevalence.

\subsection{Quantifying measure bias}
Since both RR and $\phi$ correlation show bias, we want to compare their magnitude of bias. This requires a scheme to correctly identify and quantify the bias for any given measure. It has long been recognized in clinical and observational studies and in epidemiology that null hypothesis tests for associations are of only limited use \cite{Fidler2005,Snyder2014}. Specifically, the rejection of the null-hypothesis and the significance level at which it is rejected does not signify the degree of association. Hence effect-sizes are necessary where the strength of association is important \cite{Cohen1992} as is likely true in most realistic cases.  \medskip

Selection based on minimum effect-size is equivalent to setting a threshold (assuming everything else to be fixed) for the measure.  We want to characterize bias in terms of its impact on selection. For definiteness, assume that we are interested in the subset of pairs comprising the top $f$ fraction of $\mathcal{D}$ (as determined by a given measure). We can equivalently express this in terms of a threshold ($\theta^{(f)}$) where all pairs with measure values greater than the threshold are selected. If the same threshold is then independently applied to the pairs in every partition for the bias variable, a fraction $f_j$ of pairs within each subset is selected. If there is no bias, then these fractions would be identical; conversely the extent of variation of the fractions across the partitions will be treated as a proxy for the bias.  \medskip

\begin{figure}[!p]
\hspace{-1cm}
\includegraphics[scale=0.28]{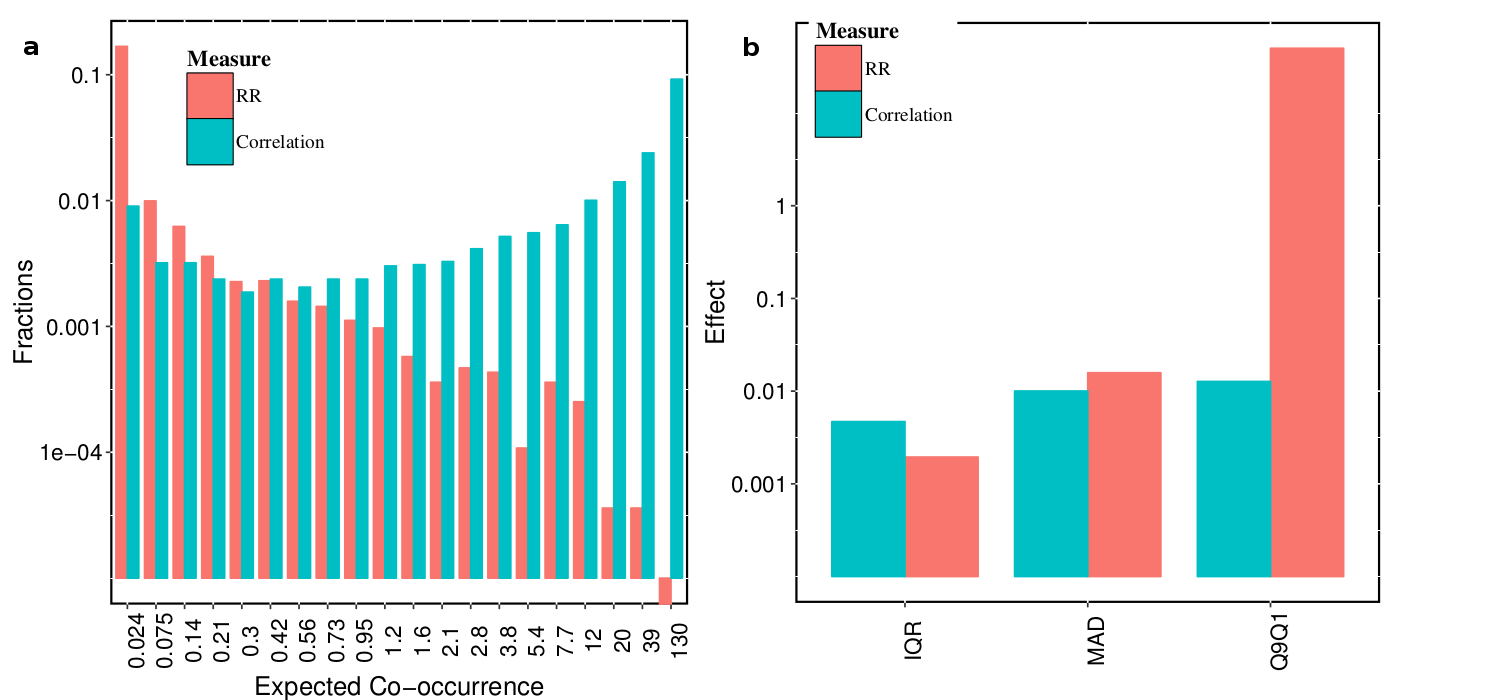}
\caption{(a) Fraction of pairs above a certain fixed threshold (corresponding to 1\% selection among all pairs) within each partition of the bias variable (expected co-occurrences) for RR and $\phi$ correlation measures. (b) Characterizing the variation of fractions across the different partitions using three indices.}
\label{BiasFrac-ExpCC}
\end{figure}

Fig. \ref{BiasFrac-ExpCC}a shows the selected fractions for RR and $\phi$ correlation measures when the overall fraction $f$ sought is 0.01 (i.e, we set a threshold such that only 1\% of all pairs are greater than that threshold) with expected co-occurrences as the bias variable. As expected the stringency of the selection depends on the partition, with larger fractions being chosen for lower expected co-occurrences in the case of RR (and the opposite for $\phi$ correlation). \medskip

This approach of using fractions to understand bias has a distinct advantage of treating all measures on an equal footing. The fractions represent the effect of the measure on the result of querying the data-set. To capture the extent of variation of fractions across the different partitions, with emphasis on its impact, we define a set of indices to characterize them. The naive approach of using standard deviation is unsatisfactory  because these fractions are typically not normally distributed. We propose three different indices that helps us better understand the bias. \medskip

1) {\bf Interquartile range (IQR)}: IQR measures difference in the location of the 75th and 25th quartile of the distribution of fractions. IQR characterizes the range within which the half of the data around the median is located. \medskip 

2){\bf  Mean Absolute Deviation (MAD)}: This weighs all deviations from the mean equally. MAD is preferred over the mean squared deviation because the underlying distribution need not be normal. \medskip

2){\bf  Q9Q1 }: To capture the most severe effect of the variation of the fractions across the bias variable, we define Q9Q1 as $\displaystyle \frac{q_{0.9} (\{f \})}{q_{0.1} (\{f \})}$, where $q_x (B)$ represents the $100x$th quantile of $B$.  This ratio of fractions (identified as significant) that are in 90th and 10th quartile emphasizes the extreme effect of the bias on comparisons.\medskip

The result of applying the three indices to the fractions shown in Fig. \ref{BiasFrac-ExpCC}a is shown in Fig. \ref{BiasFrac-ExpCC}b. We find that the Q9Q1 score has a largest gap for the two measures, and RR's bias is indeed very high (note the y-axis is logarithmic), suggesting that the effect of the bias on RR is more severe compared to $\phi$ correlation, when comparing two disease pairs whose expected co-occurrences differ widely. There is not much separating the two measures in terms of IQR or MAD. \medskip

To appreciate that our approach to studying bias is methodologically sound, we must view the partitioning and the resulting fractions as an approximation to the function $h_{Th}(\theta)$ representing the fraction of pairs selected for the bias variable $\theta$ with $Th$ being a specific threshold. If $q(\theta)$ represent the density of pairs in the population corresponding to $\theta$, then in the continuum limit,  the three indices would be defined as: \\

(i) IQR$^c$ = $h_{Th} (\theta_{0.75}) - h_{Th}(\theta_{0.25})$ where $\theta_x$ is given by the solution to the integral:
     $$ \displaystyle \int_{0}^{\theta_x} q(\theta) d\theta = x$$ \medskip
(ii) MAD$^c$ = $\int q(\theta) |h_{Th}(\theta) -\bar{h}_{Th}| d \theta$ where $\bar{h}_{Th} $ is the mean.\\ \medskip
(iii) Q9Q1$^c$ = $\displaystyle \frac{h_{Th} (\theta_{0.9})}{h_{Th} (\theta_{0.1})}$ \medskip

The definitions of these indices based on finite fractions can then be seen as straightforward discretized approximations to their continuum definitions. 

\subsubsection{Prevalence ratio bias}
\begin{figure}[!p]
\hspace{-1cm}
\includegraphics[scale=0.28]{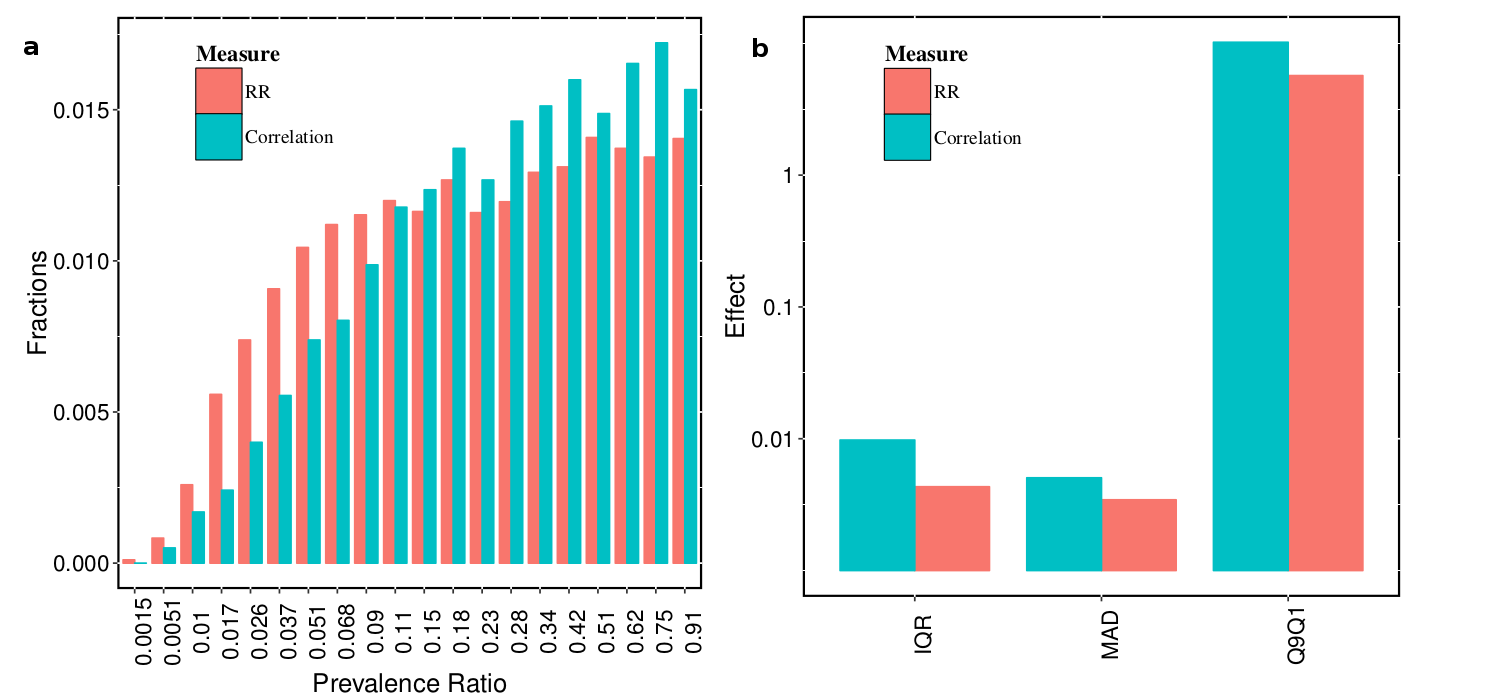}
\caption{Fraction of pairs above a certain fixed threshold (corresponding to 1\% selection among all pairs) within each partition of the bias variable (prevalence ratio, $\frac{n_{<}}{n_{>}}$, where $n_{<} (n_{>})$ is the prevalence of the less (more) frequent disease) for RR and $\phi$ correlation measures. (b) Characterizing the variation of fractions across the different partitions using three indices.}
\label{BiasFrac-PrevRatio}
\end{figure}

The ratio of the prevalences of the two diseases as a possible bias for measures has been considered before \cite{Byrt1993,Olivier2013}. We follow a similar approach as with expected-co-occurrence and partition all pairs based on $W=\frac{n_<}{n_>}$, where $n_< (n_>)$ represents the prevalence of the less (more) prevalent disease in the pair. We use the same threshold on RR and $\phi$ such that 1\% of all pairs are selected overall. The variation of the fractions is shown in Fig. \ref{BiasFrac-PrevRatio}a and we find that $\phi$ correlation shows larger differences in fractions, with distinctly suppressed numbers for low prevalence ratios. RR is relatively more balanced and this fact is also reflected in the three indices shown in Fig. \ref{BiasFrac-PrevRatio}b.

\subsubsection{Modified $\phi$ correlation}

%

The bias of $\phi$ correlation with respect to the prevalence ratio can be explained by noting that $\phi$ has an upper bound that depends on the ratio of prevalences. Assuming disease pairs with prevalence $n_A < n_B$ 
\begin{align*}
\phi(1,2) &= \frac{n_{AB}/N - (n_A/N)(n_B/N)}{\sqrt{(n_A/N)(n_B/N)(1-n_A/N)(1-n_B/N)}} \\
          &<=  \frac{\min \{ n_A, n_B \} /N - (n_A/N)(n_B/N)}{\sqrt{(n_A/N)(n_B/N)(1-n_A/N)(1-n_B/N)}} \\
          &= \frac{ (n_A/N)(1 - n_B/N)}{\sqrt{(n_A/N)(n_B/N)(1-n_A/N)(1-n_B/N)}} \\
          &= \sqrt{\frac{n_A}{n_B} \frac{1-n_B/N}{1-n_A/N}} \\
          &< \sqrt{\frac{n_A}{n_B}}
\end{align*}

Thus the maximum possible association between two diseases is not a constant but depends on their prevalence ratio \cite{Olivier2013}. This suggests that disease pairs with a high disparity in prevalences would systematically have lower values of $\phi$ correlation and indeed this is what we observed in Fig. \ref{BiasFrac-PrevRatio}a. \medskip

\begin{figure}[!p]
\hspace{-1cm}
\includegraphics[scale=0.37]{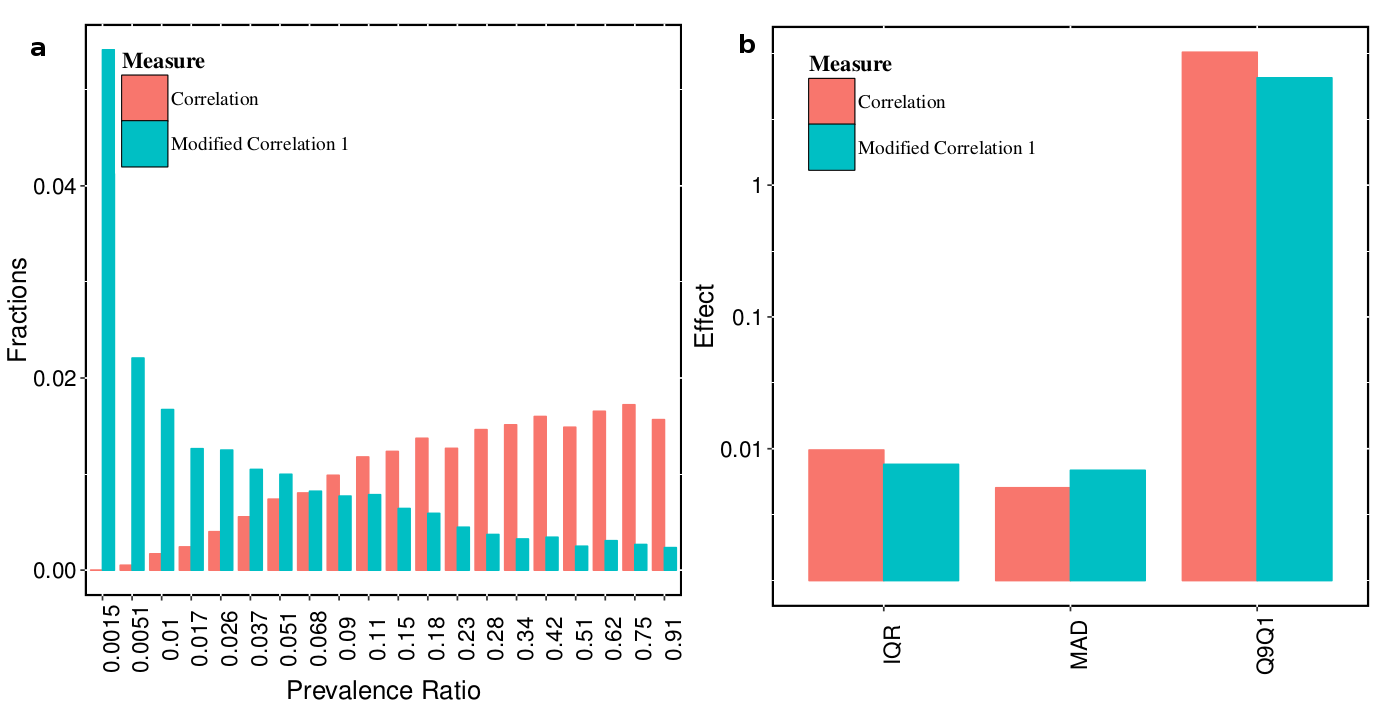}
\caption{(a) Variation of fractions with prevalence ratios for the original correlation $\phi$ and its modification $\phi^M$. We find that
the $\phi^M$ has its own bias inflating the significant fractions for smaller prevalence ratios, which is the opposite of the original measure. (b) The variation of fractions characterized by the three indices.}
\label{BiasFrac-Corr-0-1-PrevRatio}
\end{figure}

A quick workaround  of this problem is defining a modified $\phi^M = \frac{n_{AB}/N - (n_A/N)(n_B/N)}{\min \{ n_A/N, n_B/N \} } $, which has a uniform upper bound of unity, attained when co-occurrence $n_{AB} = \min \{ n_A,n_B \}$. Fig. \ref{BiasFrac-Corr-0-1-PrevRatio}a compares the original correlation and the new modified version. We immediately observe the correction offered by $\phi^M$ to the original correlation goes past the required bias removal: the measure shows a bias in the opposite direction, tending to select greater fraction for more dissimilar prevalence ratios \cite{Davenport1991}. The three indices in Fig. \ref{BiasFrac-Corr-0-1-PrevRatio}b shows that $\phi^M$ in fact has higher bias than $\phi$.  \medskip

%
%
%
%

\section{Novel measure}
We propose a conceptual and systematic approach to define a new measure of association and leading to a family parametrized by a constant $\gamma$  
\begin{equation}
 \phi_{\gamma} = \left( \frac{n_1 n_2}{N} \right) \frac{\frac{n_{12}}{N} - \frac{n_1 n_2}{N^2}}{\left(\frac{n_1+n_2}{N} \right)^2  - \left( \frac{n_1 -n_2}{N} \right)^2 \gamma}.
 \label{NM-Gamma}
\end{equation}
The full derivation of this measure (Eq. \ref{NM-GF}) is given in the Methods section but the approach and motivation is as follows. We formulate the association between a pair of diseases in terms of the differential  rate of development of one disease in the presence or absence of the other. We then use stochastic differential equation to evolve the disease probabilities in time and use the contingency table entries as constrains to estimate these rates. Since these rates are {\it a priori} unknown, and there are more variables than equations, we further use an additional constraint to choose a unique solution (the parameter $\gamma$ originates from the constraint equation). \medskip

\begin{figure}[!p]
\hspace{-1cm}
\includegraphics[scale=0.37]{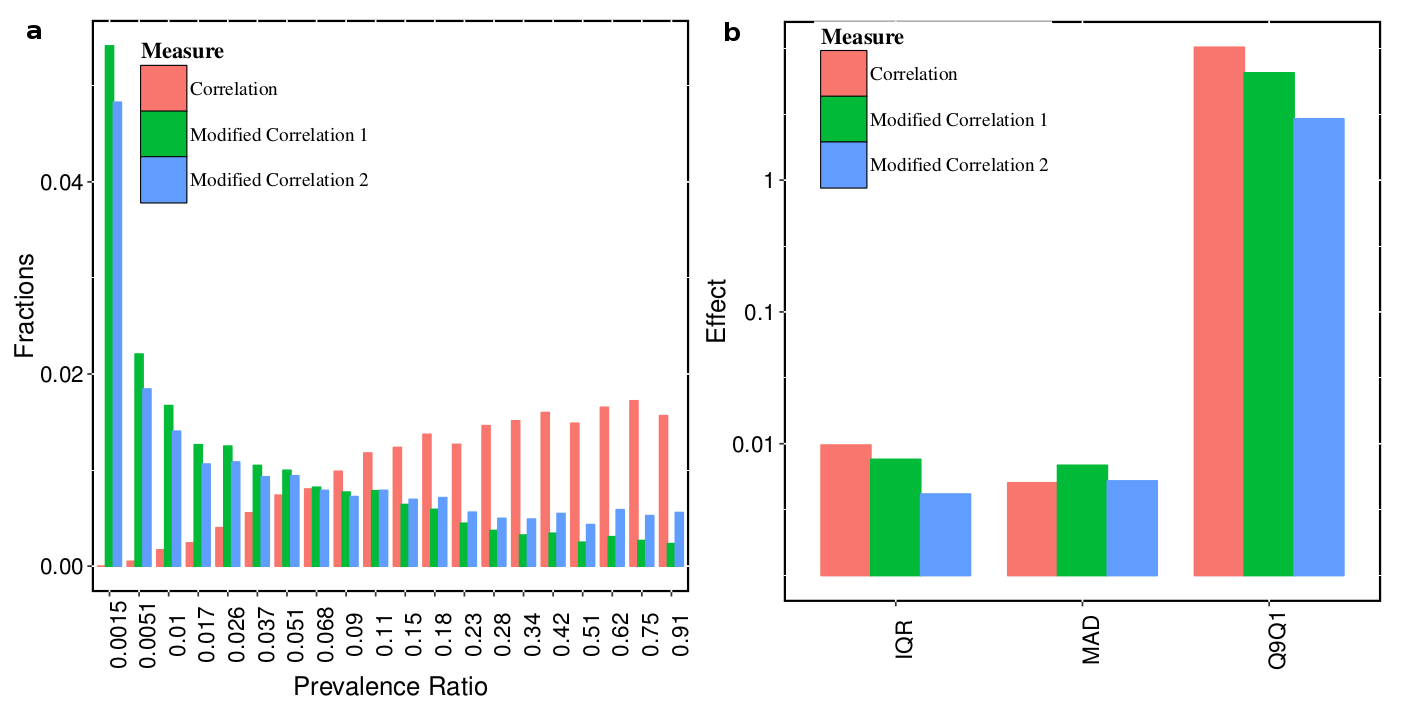}
\caption{(a)Variation of fractions with prevalence ratios across $\phi$, $\phi^M$ and $\phi^{M2}$. (b) The same is characterized by the three indices.}
\label{FracMetrics-Corr-0-1-2-PrevRatio}
\end{figure}

Defining $\phi^{M2}$ as the measure corresponding to $\gamma =1$: 
\begin{equation}
\phi^{M2} = \phi_{\gamma=1.0} = \left( \frac{n_1 +n_2}{N} \right) \frac{\frac{n_{12}}{N} - \frac{n_1 n_2}{N^2}}{  4\frac{n_1 n_2}{N^2}}
\label{Phi-M3}
\end{equation}
The justification of the choice for $\gamma$ is given in the Supplement, along with a discussion of the measure properties.  The fractions obtained when pairs are partitioned by prevalence ratio for $\phi,\phi^M$ and $\phi^{M2}$ are compared in Fig. \ref{FracMetrics-Corr-0-1-2-PrevRatio}a. Visual inspection suggests that $\phi^{M2}$ is the most balanced among them.  The corresponding indices in Fig. \ref{FracMetrics-Corr-0-1-2-PrevRatio}b confirm our observations that $\phi^{M2}$  has the least bias for all three indices, and while standard correlation performs comparably well with MAD, $\phi^{M2}$ is significantly better than both the other measures with IQR and Q9Q1.

\subsection{Comparison across threshold fractions}
We have thus far demonstrated that the new measure has the least bias with respect to prevalence ratio compared to other correlation measures. However, this was done for a specific setting of threshold, such that an overall 1\% of pairs are selected.  The next step is to determine if that reduction in bias is valid for a wider range of fractions. Indeed, the threshold setting for determining significant pairs would depend on the context of the inquiry. \medskip

A comprehensive comparison of the four measures (RR, $\phi$, $\phi^M$ and $\phi^{M2}$) is done across five different thresholds (corresponding to overall fraction of selected pairs, 0.1,0.05,0.01,0.005 and 0.001). The basis of comparison is the three indices (IQR, MAD, Q9Q1), taken one at a time. \medskip

\begin{figure}[!p]
\hspace{-1cm}
\includegraphics[scale=0.45]{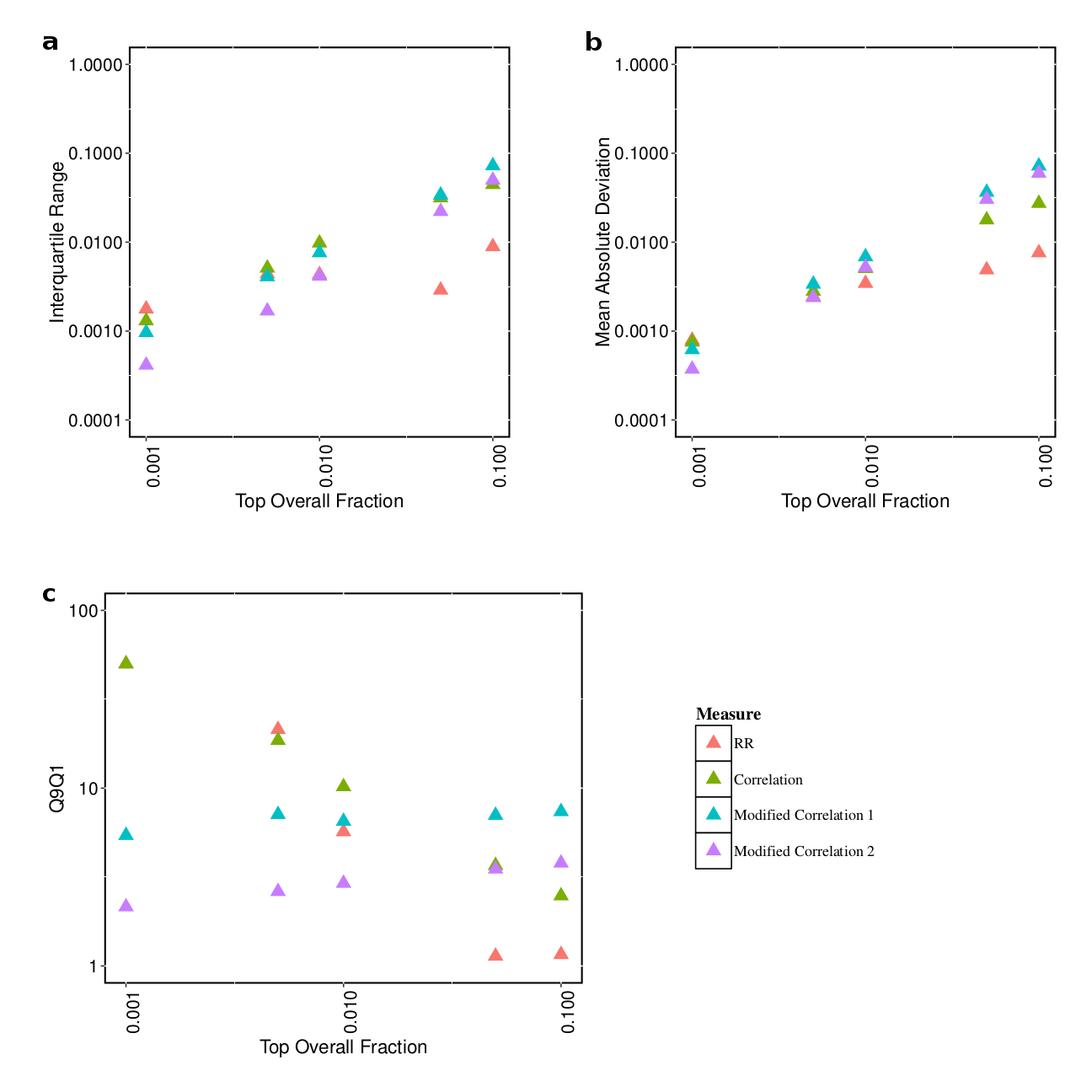}
\caption{The three indices (a) Interquartile Range (b) Mean Absolute Deviation and (c)Q9Q1 of the fractions obtained for the bias with respect to the prevalence ratio, for different overall (full data) fractions (x-axis) and for all the measures. }
\label{3Ind-PR}
\end{figure}

Fig. \ref{3Ind-PR}a,b and c shows the relative performance of all measures according to IQR, MAD and Q9Q1 respectively when bias variable is the prevalence ratio. We find, for example, from the first two sublots for IQR and MAD, that $\phi^{M2}$ has the least bias for threshold fractions less than or equal to 0.01 but has significantly more bias than RR for higher fraction of selections. For Q9Q1, $\phi^{M2}$ has the least bias for overall fractions less than or equal to 0.001. At the other end, either $\phi$ or $\phi^M$ have the highest bias for any given overall fraction and any given index. \medskip

It is clear that, although the new measure $\phi^{M2}$ is consistently better than the two correlation measures, RR has lower bias when the selected fractions are higher. There is nothing unexpected about this because the new measure was devised to eliminate the bias in $\phi$ and $\phi^M$ only.  While it may be tempting to go with RR for higher threshold fractions, we cannot prematurely conclude that until we consider the bias due to the expected co-occurrence as well. \medskip

\begin{figure}[!p]
\hspace{-1cm}
\includegraphics[scale=0.42]{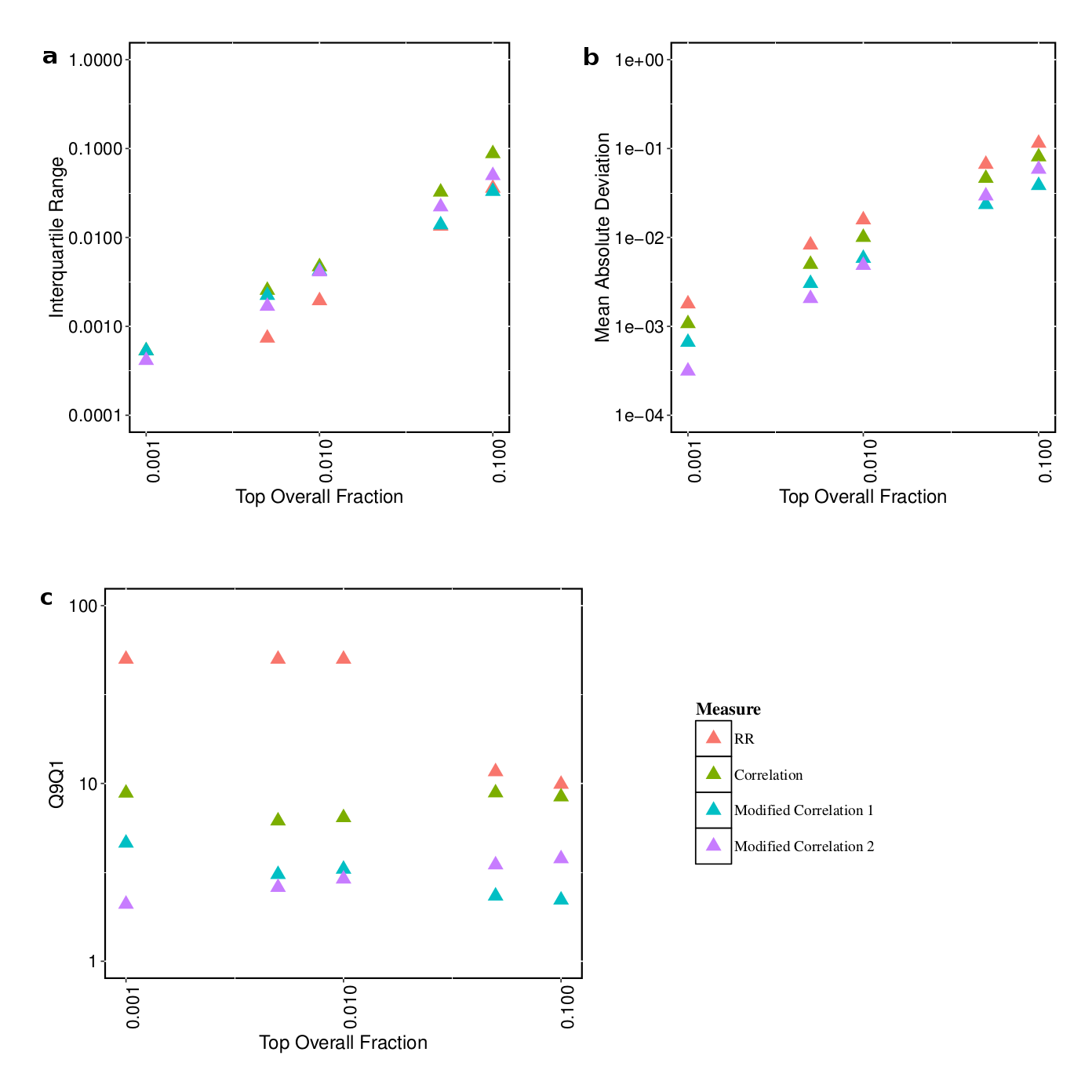}
\caption{The three indices (a) Interquartile Range (b) Mean Absolute Deviation and (c)MinMax of the fractions obtained for the bias with respect to the expected co-occurrences, for different overall (full data) fractions (x-axis) and for all the measures.}
\label{3Ind-ExpCC}
\end{figure}

Figures \ref{3Ind-ExpCC}a,b and c explore the biases with respect to the expected co-occurrence using IQR, MAD and Q9Q1 indices respectively. We find that, although we had observed in Fig. \ref{3Ind-PR} that  RR performed well for overall fractions greater than or equal to 0.05, this is not the case when we examine the bias due to expected co-occurrence. Both MAD and Q9Q1 indices show RR having the highest bias for these thresholds. \medskip

Although the different indices characterize different aspects of the bias (rankings of measures by a pair of indices need not be concordant as we can see in the above figures), it is still useful to define a combined score based on all the three indices and both the bias variables. We define such a score $s$ as the net bias  
\begin{equation}
s = (IQR)_E (MAD)_E(Q9Q1)_E + (IQR)_P (MAD)_P(Q9Q1)_P
\label{S-Score-Eqn}
\end{equation}

where the subscripts $E,P$ for each index represent the bias variable expected co-occurrence and prevalence ratio respectively.   \medskip

\begin{figure}[!p]
\hspace{-1cm}
\includegraphics[scale=0.7]{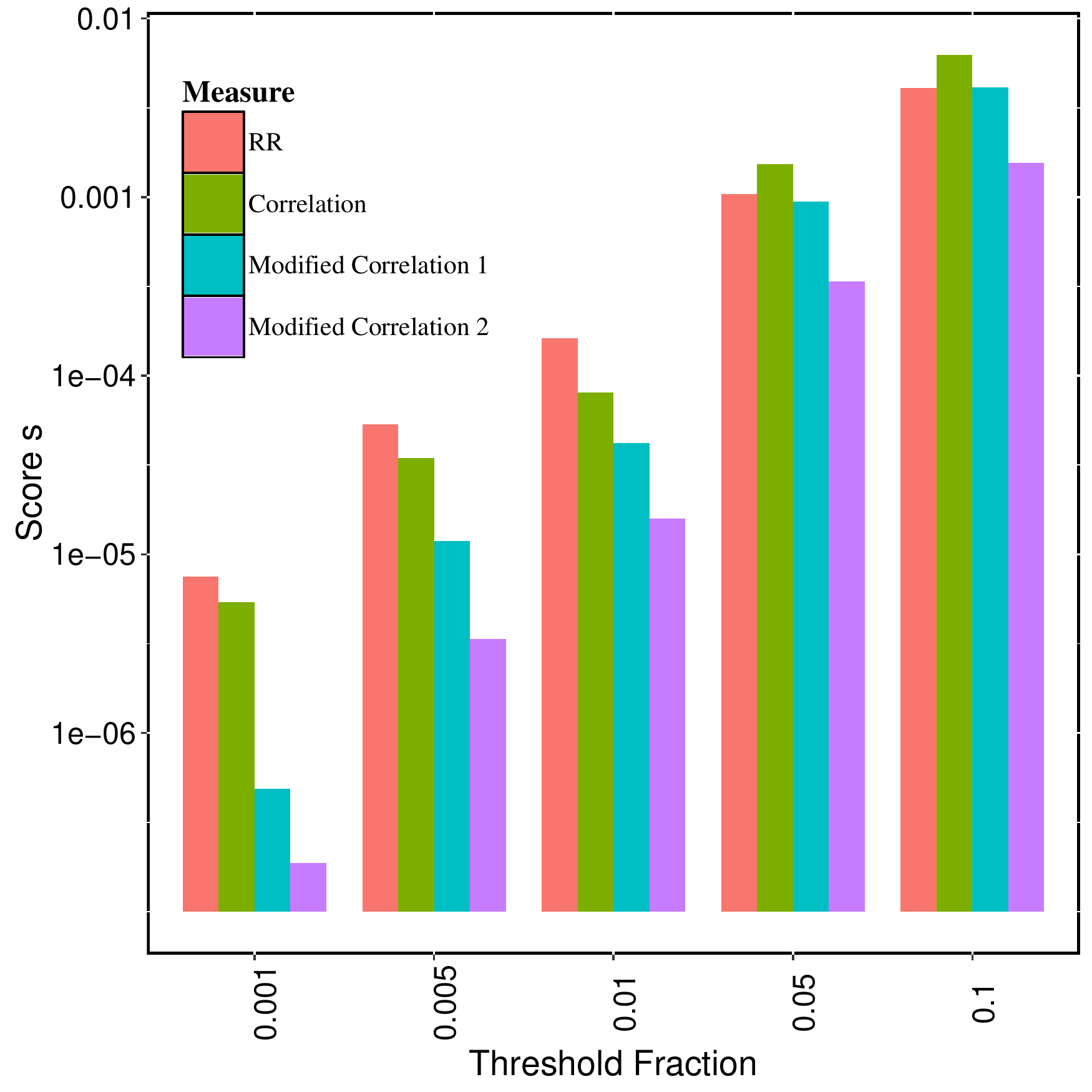}
\caption{Variation of the combined bias score s for measures RR, $\phi$, $\phi^M$ and $\phi^{M2}$ and across overall threshold fractions 0.001,0.005,0.01,0.05 and 0.01.}
\label{Score-S}
\end{figure}

Fig. \ref{Score-S} shows that, across all the thresholds considered over three orders of magnitude, $\phi^{M2}$ has the best overall performance with the least $s$ among all measures. Greater the stringency of selection, greater is the performance gain that is observed. \medskip

\begin{figure}[!p]
\hspace{-1cm}
\includegraphics[scale=0.7]{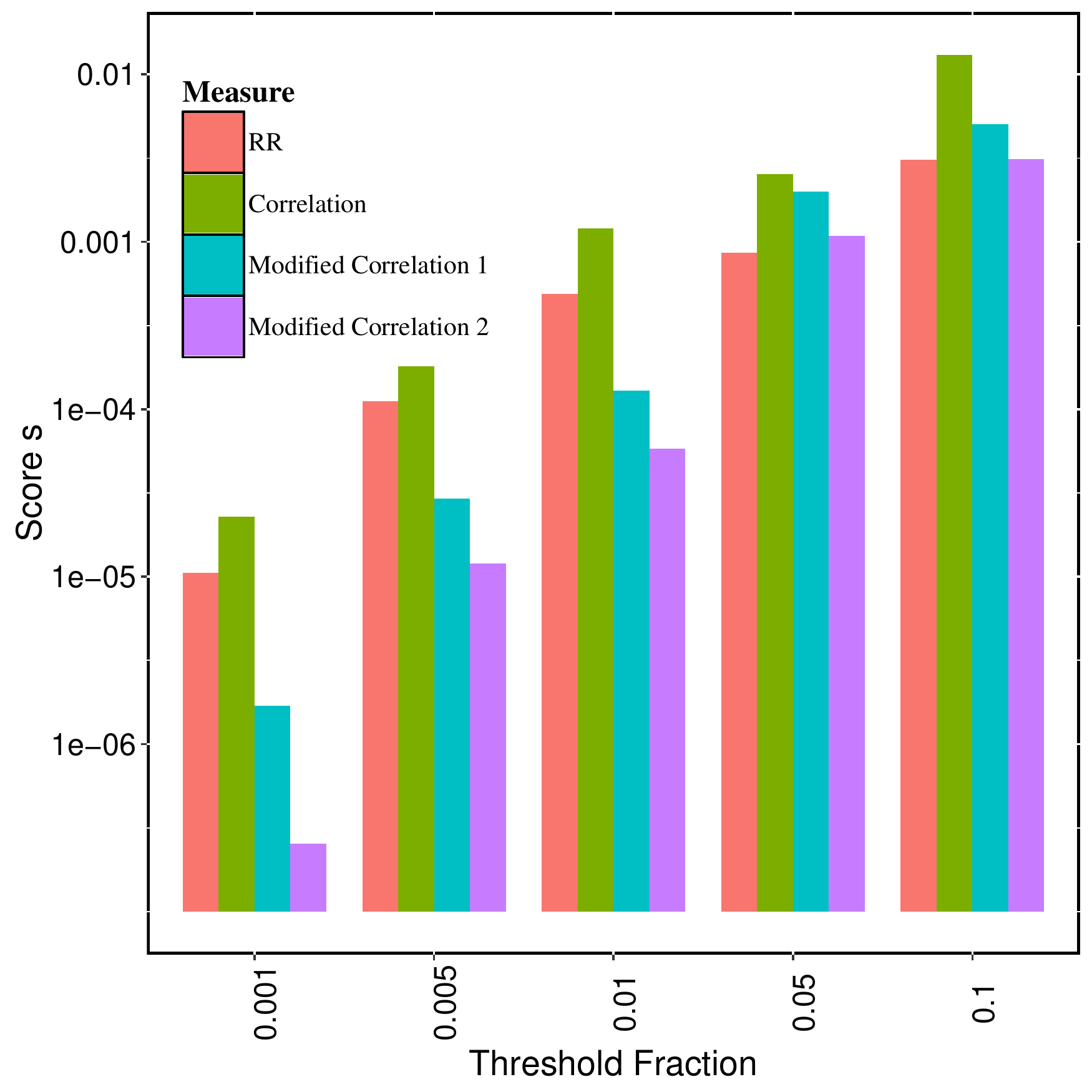}
\caption{Variation of the combined bias score s for measures RR, $\phi$, $\phi^M$ and $\phi^{M2}$ and across overall threshold fractions 0.001,0.005,0.01,0.05 and 0.01 calculated on the U.S. Medicare data.}
\label{Score-S-Med}
\end{figure}

To ensure that the bias reduction obtained with the new measure is not limited only to our cohort, we also applied these measures to another large publicly available dataset. The cohort data consists of inpatient claims of 32 million individuals enrolled in the U.S. Medicare program between 1990 and 1993, and this dataset is used in one of the early publications on disease networks \cite{Hidalgo2009}, and is now freely available. We repeated the entire procedure described above on that data set, and once again we observed that $\phi^{M2}$ had the least overall bias. Fig. \ref{Score-S-Med} shows the $s$ score for all measures across all thresholds for this dataset. We find that, while $\phi^{M2}$ performs better than all the other correlation measures irrespective of thresholds, its bias is similar to that of RR for threshold fractions 0.05 and 0.1. Nevertheless, the superior performance with other thresholds gives it the preference over RR.

\section{Methods}
\label{Meth} 

Our approach to the new measure starts with consideration of the relative probabilities to develop one disease following another. Let $\eta_1(t)$ and $\eta_2 (t))$ be boolean random variables corresponding to the two diseases (note the slight change in notation from earlier) which take values $1(0)$ when the disease is present (absent) at time t. We want to obtain a set of relations between the probabilities of occurrence and co-occurrence of the two diseases in the population at the end of time $\tau$ assuming that neither disease was present at start. To that end, we assume that the probability to be diagnosed with the disease is given by a Poisson process. However, we assign different, {\it a priori} unknown, rates to the Poisson process of a given disease depending on whether or not the other disease has already been diagnosed. For example, a given realization would be the following: starting from being disease free, disease 1 is contracted at time $t_a$ following which the Poisson rate for contracting disease 2 is different.  \\

More specifically, if $t_{E1} (t_{E2})$ represent the time point when disease $1(2)$ was diagnosed, 

\begin{align}
P( t<t_{E1} < t+ \delta t | \eta_2(t) =0)    &= \lambda_{1P} \delta t \\
P( t<t_{E1} < t+ \delta t | \eta_2(t) =1)    &= \lambda_{1S} \delta t 
\end{align} 

In the first case, the primary rate $\lambda_{1P}$ determines development of disease 1 in the absence of disease 2, but if disease 2 has been contracted before, then there is the secondary rate $\lambda_{1S}$. Likewise, the rate determining development of disease 2 before (after) diagnosis of disease 1 is given by $\lambda_{2P} (\lambda_{1S})$. \medskip

The conditional probabilities at finite time $t$:

\begin{align*}
P ( \eta_1 (t) =1 | \eta_2 (t) =0) &=  \nn \\
     \frac{ P(\eta_1 (t) =1 , \eta_2 (t) =0) }{P (\eta_2 (t) =0 )}   &= \nn \\ 
     \frac{\int_{0}^{t} P(\eta_1(t') = 0 , \eta_2 (t') = 0)  P (t' < t_{E1} < t' + \delta t' ) P (\eta_2 (t'')=0| t' < t'' < t ) dt'}{P (\eta_2 (t) =0 )}&=  \nn \\  
                                    \frac{ \int_0^{t} e^{- \lambda_{1P} t'} e^{- \lambda_{2P} t'} \lambda_{1P} \delta t' e^{- \lambda_{2S} (t-t')} } {P (\eta_2 (t) =0 ) }  &= \nn \\
                                   \frac{e^{- \lambda_{2S} t} - e^{- (\lambda_{1P} + \lambda_{2P}) t} }{P (\eta_2 (t) =0 ) (\lambda_{1P} + \lambda_{2P} -\lambda_{2S} )} 
\end{align*}

At the end time point $\tau$ : 
\begin{align}
P(\eta_1 (\tau) =1 ) &= P( \eta_1 (\tau) =1 | \eta_2 (\tau) =0) P(\eta_2 (\tau) =0) + P (\eta_1 (\tau) =1, \eta_2 (\tau) =1)  \nn\\
                     &= \frac{e^{- \lambda_{2S} t} - e^{- (\lambda_{1P} + \lambda_{2P}) t} }{(\lambda_{1P} + \lambda_{2P} -\lambda_{2S} )}  + n_{12}/N \nn \\
\label{P1-exp}                     &= n_1/N
\end{align}
where we identify the conditional probabilities at the end point with empirical values from the data :
$  P (\eta_1 (\tau) =1, \eta_2 (\tau) =1) = n_{12}/N$ and $   P (\eta_1 (\tau) =1 ) = n_1/N$

Likewise, for disease 2:
\begin{align}
P(\eta_2 (\tau) =1 ) &= P( \eta_2 (\tau) =1 | \eta_1 (\tau) =0) P(\eta_1 (\tau) =0) + P (\eta_2 (\tau) =1, \eta_1 (\tau) =1)  \nn\\
                     &= \frac{e^{- \lambda_{1S} t} - e^{- (\lambda_{1P} + \lambda_{2P}) t} }{(\lambda_{1P} + \lambda_{2P} -\lambda_{1S} )}  + n_{12}/N \nn \\
                     &= n_2/N
\label{P2-exp}
\end{align}

We can write the probability of co-occurrence $  P (\eta_1 (\tau) =1, \eta_2 (\tau) =1)$ as a sum of two probabilities for two mutually exclusive sets of events, one where disease 1 precedes disease 2, and second where this order is reversed. 
\begin{align}
P(\eta_1 (\tau) =1, \eta_2 (\tau) =1) &=  \nn \\ \int_{0}^{\tau} P(\eta_1 (\tau) = 1 | \eta_1 (t_{E2}) = 0 )  P (\eta_1 (t_{E2}) =0, t(E2) =t_{E2}) dt_{E2}  
  &+   \nn \\ \int_{0}^{\tau} P(\eta_2 (\tau) = 1 | t(E1)=t_{E1}, \eta_2 (t_{E1}) = 0 )  P (\eta_2 (t_{E1}) =0, t(E1) =t_{E1}) dt_{E1}
\label{JP}
\end{align}
where the first(second) factors accounts for cases where diagnosis of disease 1 was made after (before) disease 2.  The factors in the integrand above are : 

\begin{align*}
P(\eta_1(\tau)=1| t(E_2)=t_{E_2}, \eta_1(t_{E_2})=0) &=  1 - e^{ - \lambda_{1S} (\tau -t_{E_2})} \\ 
P(\eta_2(\tau)=1| t(E_1)=t_{E_1}, \eta_1(t_{E_1})=0) &=  1 - e^{ - \lambda_{2S} (\tau -t_{E_1})}
\end{align*}
and similarly:
\begin{align*}
P(\eta_1 (t_{E_2})=0, t(E_2) = t_{E_2}) dt_{E_2}  &= e^{-\lambda_{1P} t_{E_2}} e^{- \lambda_{2P} t_{E_2}} \lambda_{2P} dt_{E_2} \\
P(\eta_1 (t_{E_1})=0, t(E_1) = t_{E_1}) dt_{E_1}  &= e^{-\lambda_{2P} t_{E_1}} e^{- \lambda_{1P} t_{E_1}} \lambda_{1P} dt_{E_1}
\end{align*}

Plugging this in Eq. (\ref{JP}), the first integral becomes: 
\begin{align*}
\int_{0}^{\tau} & P(\eta_1 (\tau) = 1 | t(E_2)=t_{E_2}, \eta_1 (t_{E_2}) = 0 )  P (\eta_1 (t_{E_2}) =0, t(E_2) =t_{E_2}) dt_{E_2} \\
  &= \int_{0}^{\tau} (1 -e^{-\lambda_{1S} (\tau -t_{E_2})}) e^{-\lambda_{1P} t_{E_2}} e^{-\lambda_{2P} t_{E_2}} \lambda_{2P} dt_{E_2} \\
  &= \frac{\lambda_{2P}}{\lambda_{1P} + \lambda_{2P}} (1-e^{- (\lambda_{1P} + \lambda_{2P}) \tau} )  - e^{- \lambda_{2S} \tau} \frac{\lambda_{2P}}{\lambda_{2P} +\lambda_{1P} - \lambda_{1S}} (1 - e^{-(\lambda_{1P} + \lambda_{2P} - \lambda_{2S})\tau})    
\end{align*}
If we make the approximation $ \lambda_{i \alpha} \tau<<1$ (W1)  for $i=1,2$ and $\alpha = P,S $, then the above reduces to $\lambda_{1S} \lambda_{2P} \tau^2 /2$. 
Correspondingly we for the the second integral in Eq. (\ref{JP}) the approximation $\lambda_{1P} \lambda_{2S} \tau^2 /2$. 
leading to:
\begin{equation} 
\frac{n_{12}}{N} = \frac{(\lambda_{1P} \lambda_{2S} + \lambda_{1S} \lambda_{2P}) \tau^2}{2}
\label{A1-CC-Rel}
\end{equation}

Under the same assumption W1, we have:
$$\frac{e^{- \lambda_{2S} \tau} - e^{- (\lambda_{1P} + \lambda_{2P}) \tau}}{(\lambda_{1P} + \lambda_{2P} -\lambda_{2S} )} \sim \tau - (\lambda_{1P} + \lambda_{2P} -\lambda_{2S})\tau^2$$ 
$$\frac{e^{- \lambda_{1S} \tau} - e^{- (\lambda_{1P} + \lambda_{2P}) \tau}}{(\lambda_{1P} + \lambda_{2P} -\lambda_{1S} )} \sim \tau - (\lambda_{1P} + \lambda_{2P} -\lambda_{1S})\tau^2$$

Plugging the above in Eq. (\ref{P1-exp},\ref{P2-exp}), we obtain :
\begin{subequations}
\label{P1P2}
\begin{align}
n_1/N &= (\lambda_{1P} \tau) - \lambda_{1P} \frac{\lambda_{1P} + \lambda_{2P}}{2} \tau^2 - \lambda_{2P} \lambda_{1S} \tau^2  \\
n_2/N &= (\lambda_{2P} \tau) - \lambda_{2P} \frac{\lambda_{1P} + \lambda_{2P}}{2} \tau^2 - \lambda_{1P} \lambda_{2S} \tau^2  
\end{align}
\end{subequations}
Eqs. (\ref{A1-CC-Rel}, \ref{P1P2}) are a set of three equations but with four unknowns $ \lambda_{i \alpha},\, i=1,2 \qquad \alpha =P,S$, which we cannot solve without an additional simplifying assumption. This is of course what we would expect, because that extra degree of freedom corresponds to our ignorance of the underlying causal relations between the two diseases.  \medskip

For example if we use the following ansatz :

$$ \lambda_{iS} = \lambda_{iP} q, \qquad i=1,2$$ 

we can solve the equations and $ q = \frac{n_{12}/N}{(n_1 - n_{12}/2)/N (n_2 - n_{12}/2)/N}$, which would approximate to the standard definition of relative risk  ( $n_{12}/\min{n_1,n_2} << 1)$.   \medskip

Substituting $\lambda_{iS} = \lambda_{iP} + q_i, \qquad i=1,2$, we can rewrite the above relations in terms of $q_i$ which represents the deviation from the situation where the two diseases are unrelated, i.e, $q_i =0$.

\begin{align}
\frac{n_1}{N} &= \lambda_{1P} \tau  - \frac{(\lambda_{1P} \tau)^2}{2} + \frac{\lambda_{2P} q_1 \tau^2}{2} \nn \\
\frac{n_2}{N} &= \lambda_{2P} \tau  - \frac{(\lambda_{2P} \tau)^2}{2} + \frac{\lambda_{1P} q_2 \tau^2}{2} \nn \\
\frac{n_{12}}{N} &= \lambda_{1P} \lambda_{2P} \tau^2 + \frac{\lambda_{1P} q_2 + \lambda_{2P} q_1}{2} \tau^2 \label{CC-P}
\end{align} 
Further simplifying using A1 we have : 
\begin{subequations}
\begin{align}
\lambda_{1P} \tau &= \frac{n_1}{N} - q_1 \tau \frac{\lambda_{2P} \tau}{2} \label{FPR1} \\
\lambda_{2P} \tau &= \frac{n_2}{N} - q_2 \tau \frac{\lambda_{1P} \tau}{2} \label{FPR2}
\end{align}
\end{subequations}
and plugging into Eq. (\ref{CC-P}) 
\begin{align}
n_{12}/N &=  \nn \\ (n_1/N - \frac{q_1 \tau n_2}{N})(n_2/N - \frac{q_2 \tau n_1}{2})  +  \frac{(n_1/N - q_1 n_2 \tau/N)q_2 \tau}{2} + \frac{(n_2/N - q_2 n_1 \tau /N) q_1 \tau}{2})
&=  \nn \\ \frac{n_1 n_2}{N^2} ( 1 + q_1 q_2 \tau^2 ) + \frac{n_1 q_2 \tau}{2N} (1 - n_1/N) -\frac{n_2 q_1 \tau}{2N} (1 - \frac{n_2}{N}) - q_1 q_2 \tau^2 \frac{n_1 + n_2}{N}  
&\sim  \nn \\ \frac{n_1 n_2}{N^2} + \frac{n_1 q_2 \tau}{2N}  + \frac{n_2 q_1 \tau }{2N} \nn
\end{align}
         
where we have arrived at the third expression by dropping terms of order $ \frac{n_1n_2}{N^2}q_i \tau$, $\frac{n_i}{N}q_1q_2 \tau^2$ and higher. Rewriting the last step, 
\begin{equation}
\frac{n_1 q_2 \tau + n_2 q_1 \tau}{N} = 2 (\frac{n_{12}}{N} - \frac{n_1 n_2}{N^2}) 
\label{eq-dev}
\end{equation}

Eq. (\ref{eq-dev}) relates the two unknowns $q_1$ and $q_2$. As we have no other constraint apriori, there is a family of solution to this equation. We have already examined one example above. \medskip

Thus, we need another constraint in order to determine $q_i$'s. We posit two hueristic factors in this regard: one, apriori we would expect $q_i$'s to be close to one another, and the second, maximization of their sum. If we only had the second criterion, then the $q$ corresponding to be the disease with lower prevalence would be 0, and that of the higher prevalent disease very high.  Imposing the first criterion alone assumes a symmetry between the diseases, and while that may be reasonable in the absence of any other information, we instead consider a trade-off between them.  We propose the minimization of the "energy" function:

\begin{equation}
E = \alpha (q'_1 -q'_2)^2  - \beta (q'_1 +q'_2)^2
\label{Egy}
\end{equation} 
where $q'_i =q_i \tau$, $\alpha,\beta >0$; the first term favors the $q'_i$'s being close together and the second maximizing the sum. \medskip

We look for solutions of Eqs. (\ref{eq-dev})  that maximizes Eq. (\ref{Egy}). This is done by using the Lagrange multiplier technique for finding the extrema given a constraint. 
$L(q'_1,q'_2) = E + \rho (q'_1 n_2/N + q'_2 n_1/N - (n_{12}/N - n_1n_2/N^2))$
\begin{align*}
\frac{\partial L}{ \partial q'_1} &= 2 \alpha (q'_1 - q'_2)  -2 \beta (q'_1 +q'_2) + \rho n_2/N \\ 
\frac{\partial L}{\partial q'_2} &= 2 \alpha (q'_2 -q'_1)  - 2\beta (q'_1 + q'_2) + \rho n_1/N  
\end{align*}  

We can now solve the above constraint equations together with Eq. \ref{eq-dev} simultaneously for the three unknows $q'_1,q'_2$ and $\rho$. 
We give the final expressions for $q'_i$,i=1,2:
\begin{align*}
q'_1 &= 4 \left( \frac{n_1 +n_2}{N \beta} + \frac{n_1 - n_2}{N \alpha} \right) \frac{\frac{n_{12}}{N} - \frac{n_1 n_2}{N^2}}{\left(\frac{n_1+n_2}{N} \right)^2 \frac{1}{\beta} - \left( \frac{n_1 -n_2}{N} \right)^2 \frac{1}{\alpha}} \\
q'_2 &= 4 \left( \frac{n_1 +n_2}{N \beta} - \frac{n_1 - n_2}{N \alpha} \right) \frac{\frac{n_{12}}{N} - \frac{n_1 n_2}{N^2}}{\left(\frac{n_1+n_2}{N} \right)^2 \frac{1}{\beta} - \left( \frac{n_1 -n_2}{N} \right)^2 \frac{1}{\alpha}} 
\end{align*}
And the sum gives us the desired measure in terms of prevalence and co-occurrence
\begin{equation*}
q'_{tot} = q'_1 +q'_2 = 8 \left( \frac{n_1 +n_2}{N \beta} \right) \frac{\frac{n_{12}}{N} - \frac{n_1 n_2}{N^2}}{\left(\frac{n_1+n_2}{N} \right)^2 \frac{1}{\beta} - \left( \frac{n_1 -n_2}{N} \right)^2 \frac{1}{\alpha}}.
\label{q-tot-asym}
\end{equation*}
This represents a family of measures parametrized by $\alpha,\beta >0$. Defining $\gamma = \beta/\alpha$ and skipping the constant multiplicative factor of 8: 
\begin{equation}
 \phi_{\gamma} = \left( \frac{n_1 + n_2}{N} \right) \frac{\frac{n_{12}}{N} - \frac{n_1 n_2}{N^2}}{\left(\frac{n_1+n_2}{N} \right)^2  - \left( \frac{n_1 -n_2}{N} \right)^2 \gamma}.
 \label{NM-GF}
\end{equation}
 
For the symmetric case $n_1 = n_2 =n$, we have:
\begin{equation*}
q'_{tot} = 0.5 \frac{\frac{n_{12}}{N} - \frac{n^2}{N^2}}{\frac{n}{N}}
\end{equation*}
which, except for the factor of 0.5, is very close to what we would get with the original $\phi$ and modified $\phi^M$ correlation, Eq. (\ref{Phi-Org}), and exact in the limit of vanishing $n/N$. Although this result is independent of constants $\alpha, \beta$, to get reasonable answers for arbitrary ratios $n_2/n_1$, we require $\alpha/ \beta \sim 1$ (see Supplement). 

\section{Discussion}


Our proposed framework to characterize the measure properties across potential bias variables has three key features: (a) bias is characterized in terms of its impact on selection based on effect-size (b) procedure for determining bias is independent of the specific measure and hence comparison between measures is carried out on a neutral platform (c) three indices are devised to capture different aspects of the bias.  \medskip

The importance of understanding the properties of measures cannot be overstated. Even for randomized controlled trials, conclusions depend on the measure used to characterize the effect of interventions \cite{Schechtman2002}. Although not widely recognized, it is known that most standard measures of association have significant limitations and can give rise to misleading results unless they are interpreted carefully \cite{Breaugh2003}. Despite use of similar measures, the results obtained from studies using different design of experiments cannot be directly compared or combined \cite{Morris2002}. While all these issues are certainly very relevant, we should clarify that the measure bias that is highlighted in our work here has different origins, and to the best of our knowledge, there has been no earlier studies that have addressed them. \medskip

The requirement that the measure distribution be independent of prevalence ratio could be questioned in certain limits. In the extreme case, one can argue that, for a pair of diseases with very different prevalence  $n_A >> n_B$, an unbiased measure should never assign maximal association even when the co-occurrence is the highest possible, $n_{AB}=n_B$. The basis for this is the observation that perfect association implies the pair co-occur in every case (aside from errors from misclassification or finiteness of study time scale), and hence divergent prevalence would not arise in the first place. A reasonable counter-argument would point to the potential scenario of $B$ being an invariable cause of $A$. Whatever the consensus may be on this issue, most realistic situations contain very few or no such cases, and in large cohorts the relations in the generic pair follow $n_{AB} << \min \{n_A, n_B\}$ where one would expect measure distribution to be independent of prevalence ratio. \medskip
\clearpage

\begin{table}[h!]
\center
\begin{tabular}{|c|c|c|c|}
\hline 
 & $A$  & $\bar{A}$ & \\
 \hline
 $B$ & $p$ & $q$  & $n_B = p +q$\\
 \hline
 $\bar{B}$ & $r$ & $s$ & $n_{\bar{B}} = r+s$ \\
 \hline
  & $n_A =  p+r$ & $n_{\bar{A}} = q + s$ & $N= p +q+r+s$ \\ 
 \hline
\end{tabular}
\caption{Contingency table for a pair of diseases in a cohort}
\label{Cont}
\end{table}  

%


\clearpage
\appendix
\appendixpage

\section{Effect of parameter $\gamma$ on measure bias}

\begin{figure} [!h]
\includegraphics[scale=0.5]{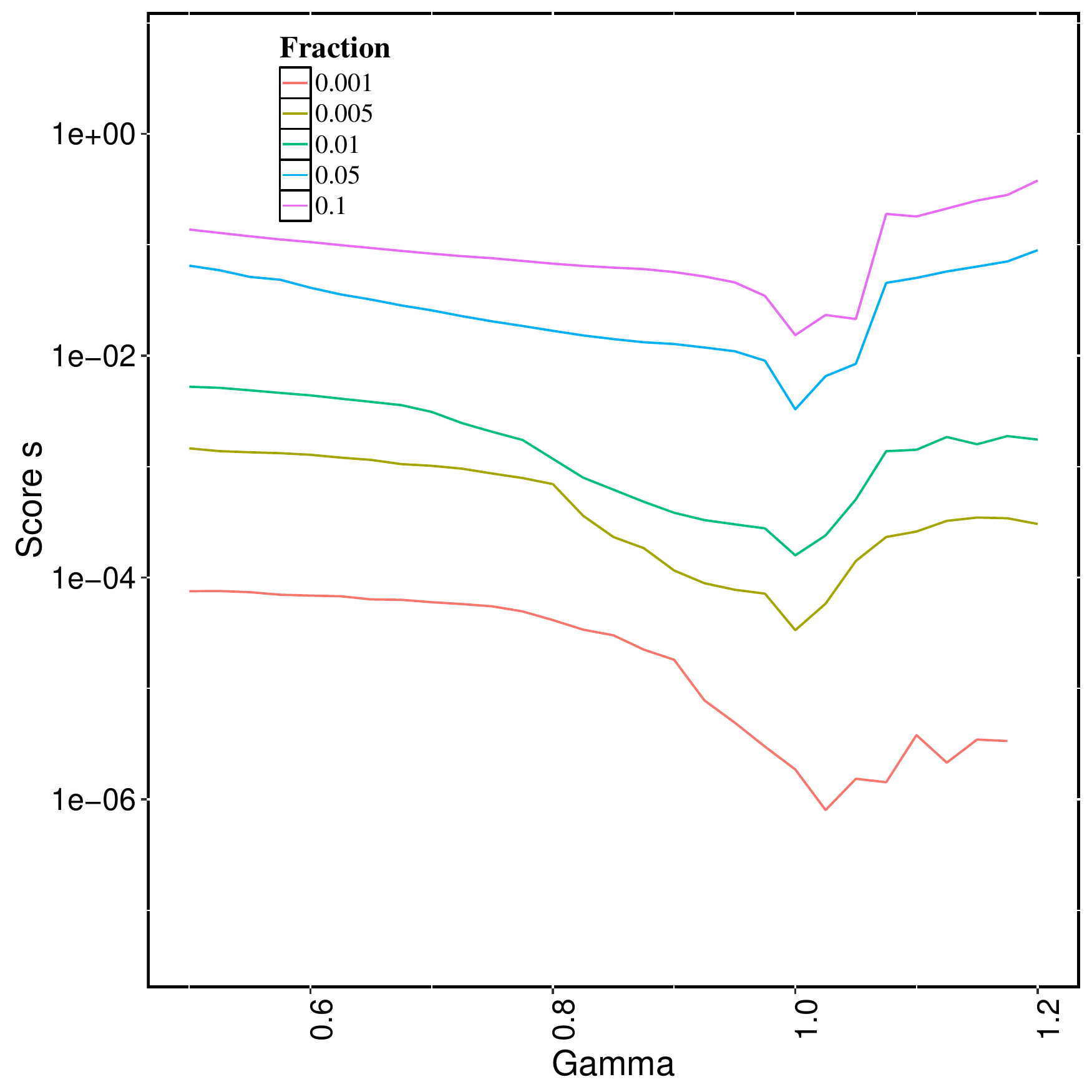}
\caption{Variation of the combined bias score $s$ of the three indices with the parameter $\gamma$ (in the new measure family) across different thresholds. }
\label{Score-S-Gamma}
\end{figure}

We consider how the measure properties change as we modify the value of the parameter $\gamma$ in Eq. 5 (main text). In particular, we are interested in the dependence of measure biases on $\gamma$. However, as we we have observed earlier, ranking of measures based on magnitude of bias defined by the three indices are not concordant.  Following our approach to compare different measures in the main text, we use the score $s$ (Eq. 7 in main text) to captures the net bias from the three indices and for both bias variables.  Fig. \ref{Score-S-Gamma} shows the variation of $s$ with $\gamma$ for the different thresholds. We find that the minimum is attained at $\gamma=1$ in almost every case. In fact, the only deviation from $\gamma=1$ occurs at threshold fraction of 0.001, and even here the minima is attained nearby at 1.025. \medskip

Besides the explicit comparisons using the score $s$, there is an intuitive reason for selecting $\gamma=1$. In the derivation of the measure, we find that $\gamma$ enters in the ``energy" equation whose minimization provides the additional constraint necessary to solve for the measure. Setting $\gamma = 1$, equivalently $\alpha = \beta$ in Eq. 19 makes the two terms symmetric. In addition, the final expression obtained has a greater intuitive appeal, being symmetric in $n_1,n_2$ and being free of any unwieldy constants. Moreover, when $n_1=n_2=n$, this is equivalent to the original $\phi$ and $\phi^M$ except for a term of order $O(n/N)$  which in most realistic cases is negligible. Also, for unequal prevalence, $n_2 < n_1$ the maximum value of the measure is
$$ \phi^ {M2}(n_{12}=n_2) = \frac{1}{4}\left(1 + \frac{n_2}{n_1} \right) \left( 1 - \frac{n_1}{N} \right).$$
Ignoring the last factor, its range is between 0.25 (in the limit $n_2 \rightarrow 0$) and 0.5 (equal prevalence). This places the measure in-between $\phi$ (maximum diminishes to 0 when $n_2 \rightarrow 0$) and $\phi^M$ (maximum is uniformly unity for all pairs of prevalence).

\clearpage

\section{Poisson Distribution of disease co-occurrence}
Here we show that distribution of co-occurrences for a given pair of diseases with specified incidence rates  under the null hypothesis that the two diseases are completely independent, is in fact, Poisson. \\

Concretely, if $N$ is total population of the cohort, and $n_i$ represents the number of people with disease $i$ ($i=1,2, \cdots M$), and if all diseases are uncorrelated with each other and occur randomly in the population, the total number of ways in which we can have an overlap of $n_{ij}$ individuals carrying both diseases $i$ and $j$ is given by:

\begin{align}
C(n_i,n_j,n_{ij},N) &= \dbinom{N}{n_i}\dbinom{n_i}{n_{ij}}\dbinom{N - n_i }{n_j -n_{ij}}    
                \notag  \\  &= \frac{N!}{(n_i -n _{ij})! (n_j -n _{ij})! n_{ij}! (N+n_{ij} -n_i -n_j)!} 
\end{align}
where the first factor corresponds to choosing $n_i$ elements (first set) from $N$, the second, the intersecting $n_{ij}$ elements among the $n_i$, and the third, the  $n_j -n_{ij}$ elements (second set - overlapping elements) from the remaining $N -n_i$ elements. 

Taking the log on both sides and using the Sterling approximation $\log{N!} = N \log{N} -N$, 

\begin{align}
\log{C(n_i,n_j,n_{ij},N)} &= N\log{N} - n_i(1-\frac{n_{ij}}{n_i})\log{n_i(1-\frac{n_{ij}}{n_i})} - n_j(1-\frac{n_{ij}}{n_j})\log{n_j(1-\frac{n_{ij}}{n_j})} \notag \\ &- n_{ij}\log{n_{ij}} - (N-n_i -n_j)(1 +\frac{n_{ij}}{N -n_i - n_j})\log{(N-n_i -n_j)(1 +\frac{n_{ij}}{N -n_i - n_j})}
\end{align}

Simplifying and retaining only terms that are first order in  $n_{ij}$ or higher:

\begin{align}
\log{C(n_i,n_j,n_{ij},N)} &= \log{C_0}  + n_{ij} ( 1 + \log{\frac{n_i n_j}{N -n_i -n_j}} -\log{n_{ij}})
\end{align}
Exponentiating, and using the Sterling approximation in reverse, we get:
\begin{equation}
C= \frac{C_0}{n_{ij}!} \left(\frac{n_i n_j}{N -n_i -n_j}\right)^{n_{ij}}
\end{equation}
which is nothing but the Poisson distribution with (average  $\mathbf{E} (n_{ij})= \lambda = \frac{n_i n_j}{N -n_i -n_j}$ and the constant $C_0$ would be normalized to $e^{-\lambda}$. \\

This is of course what we would expect from naive considerations of probability theory.  Thus one can calculate p-values from that of the Poisson distribution \\
the p-value corresponding to an observation $n_{ij}$, then comes from the cumulative distributive function of Poisson : 
\begin{equation}
F (n_{ij}) =  \displaystyle \sum_{k \geq n_{ij}}  P_{Poiss} (n_{ij},\lambda)
\end{equation}

\clearpage


\end{document}